\documentclass[fleqn,usenatbib]{mnras}

\usepackage{newtxtext,newtxmath}
\usepackage{ulem}

\usepackage[T1]{fontenc}

\DeclareRobustCommand{\VAN}[3]{#2}
\let\VANthebibliography\thebibliography
\def\thebibliography{\DeclareRobustCommand{\VAN}[3]{##3}\VANthebibliography}

\usepackage{graphicx}	
\usepackage{amsmath}	

\newcommand{\fracb}[2]{\left(\frac{#1}{#2}\right)}

\newcommand{\red}[1]{\textcolor{red}{#1}}
\newcommand{\blue}[1]{\textcolor{blue}{#1}}

\usepackage[dvipsnames]{xcolor}
\definecolor{blazeorange}{rgb}{1.0, 0.4, 0.0}
\definecolor{seagreen}{rgb}{0.18, 0.55, 0.34}
\definecolor{rufous}{rgb}{0.66, 0.11, 0.03}
\definecolor{royalfuchsia}{rgb}{0.79, 0.17, 0.57}
\definecolor{scarlet}{rgb}{1.0, 0.13, 0.0}
\definecolor{royalpurple}{rgb}{0.47, 0.32, 0.66}

\newcommand\rev[1]{{\color{black} #1}} 
\newcommand\revv[1]{{\color{black} #1}} 

\newcommand\grb{GRB\,221009A}

\title[GRB\,221009A Afterglow]{GRB\,221009A Afterglow from a Shallow Angular Structured Jet}

\author[Gill \& Granot]{
Ramandeep Gill$^{1,3}$\thanks{E-mail: r.gill@irya.unam.mx}
and Jonathan Granot$^{2,3,4}$\thanks{E-mail: granot@openu.ac.il}
\\
$^{1}$Instituto de Radioastronomía y Astrofísica, Universidad Nacional Autónoma de México, Antigua Carretera a Pátzcuaro \# 8701, \\
Ex-Hda. San José de la Huerta, Morelia, Michoacán, México C.P. 58089 \\
$^{2}$Department of Natural Sciences, The Open University of Israel, P.O Box 808, Ra'anana 4353701, Israel\\
$^{3}$Astrophysics Research Center of the Open university (ARCO), The Open University of Israel, P.O Box 808, Ra'anana 4353701, Israel\\
$^{4}$Department of Physics, The George Washington University, Washington, DC 20052, USA
}

\date{Accepted XXX. Received YYY; in original form ZZZ}

\pubyear{2023}

\begin{document}
\label{firstpage}
\pagerange{\pageref{firstpage}--\pageref{lastpage}}
\maketitle

\begin{abstract}
Exceptionally bright gamma-ray burst (GRB) afterglows can reveal the angular structure 
of their 
jets. GRB jets appear to have a narrow core (of half-opening angle 
$\theta_c$), beyond which their kinetic energy drops as a power-law with angle $\theta$ from 
the jet's symmetry axis, $E_{k,\rm iso}(\theta)\propto[1+(\theta/\theta_c)^2]^{-a/2}$. The power-law 
index $a$ reflects the amount of mixing between the shocked jet and confining medium, which depends 
on the jet's initial magnetization. Weakly magnetized jets undergo significant mixing, leading to 
shallow ($a\lesssim2$) angular profiles. We use the  exquisite multi-waveband afterglow observations 
of \grb~to constrain the jet angular structure using a dynamical model that accounts for both the 
forward and reverse shocks, for a power-law external density profile, $n_{\rm{}ext}\propto{}R^{-k}$. 
Both the forward-shock emission, that dominates the optical and X-ray flux, and the reverse-shock 
emission, that produces the radio afterglow, require a jet with a narrow core ($\theta_c\approx0.021$) 
and a shallow angular structure ($a\approx0.8$) expanding into a stellar wind ($k\approx2$). 
\revv{Moreover, these data appear to favor a small fraction ($\xi_e\approx10^{-2}$) of shock-heated electrons forming a power-law energy distribution in both shocks.}
\end{abstract}



\section{Introduction}
New, paradigm-shifting insights into gamma-ray burst (GRB) physics can be gained 
by observing exceptionally bright GRBs, such as the recent
\grb~\citep{An+23,Frederiks+23,Lesage+23,Ripa+23,Burns+23} 
that had an extremely high fluence of $S_\gamma\sim0.2\,{\rm erg\,cm}^{-2}$.  
This is a result of its relatively closer distance, with a redshift of $z=0.151$ 
\citep[e.g.][]{deUgartePostigo+22,Malesani+23}, and a record-breaking prompt $\gamma$-ray isotropic equivalent
energy release of $E_{\gamma,\rm iso}\sim1.2\times10^{55}\,$erg, over a total duration of 
$t_{\rm GRB}\sim600\,$s. 
This GRB is even more important and unique since
the Large High Altitude Air Shower Observatory (LHAASO) reported the detection of over
5000 very-high-energy photons, including an 18\,TeV photon, within 2000\,s of the burst trigger \citep{Huang+22GCN}.

The broadband afterglow of this burst was recorded in exquisite detail 
from radio to $\gamma$-rays \citep{Bright+23,Laskar+23,OConner+23,Fulton+23,Kann+23,An+23,Williams+23,Frederiks+23,Lesage+23}. 
Its modeling in some of these works and 
others \citep{Ren+23,Sato+23} with the canonical forward-shock (FS) and reverse-shock (RS) emission from a uniform and 
even a two-component jet presented some challenges. In particular, \citet{Laskar+23} found it difficult to explain the 
radio data both with the RS and FS 
afterglows in the wind ($k=2$) scenario (for an external medium density $n_{\rm ext}\propto R^{-k}$) 
which they preferred over the interstellar medium (ISM; $k=0$) case. In addition, their FS model could not adequately fit both 
the early- ($\lesssim1\,$ day) and late-time ($\gtrsim30\,$day) optical data. Some success was achieved by 
\citet{OConner+23} who modeled the FS emission from a shallow angular structured jet and the RS emission from 
a spherical outflow expanding into a uniform medium (ISM).

\begin{figure*}
    \centering
    \includegraphics[width=\textwidth]{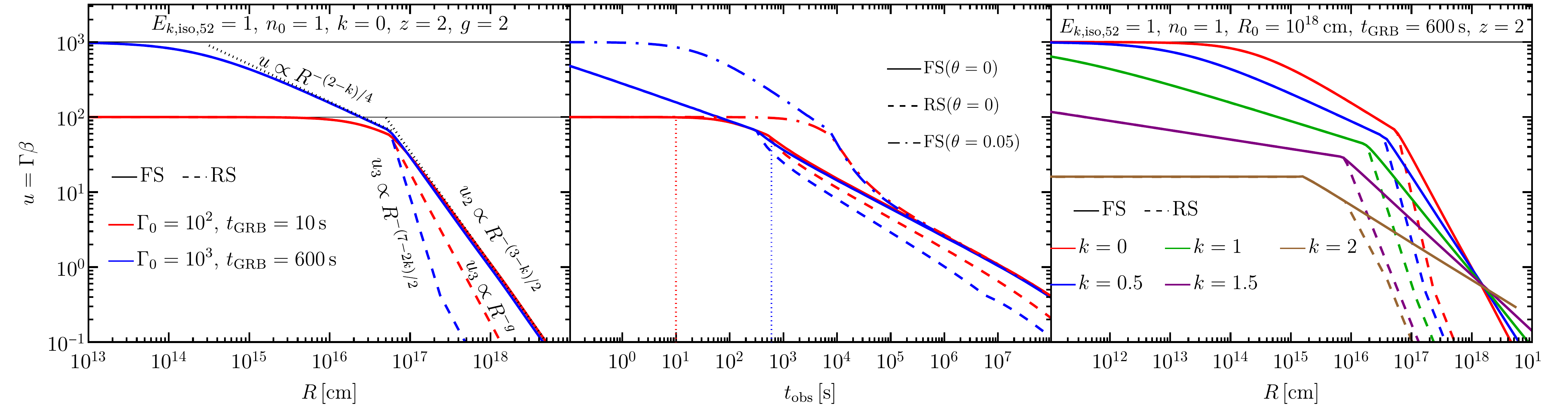}
    \vspace{-0.6cm}
    \caption{
    The proper speed $u=\Gamma\beta$ of the shocked regions downstream of the forward (FS) and reverse (RS) shocks. 
    \textbf{\textit{Left}}: Two different prompt GRB durations $t_{\rm GRB}$ and initial bulk LFs $\Gamma_0$ are chosen to obtain the thin-shell 
    (\red{red}) and thick-shell (\blue{blue}) cases. 
    \textbf{\textit{Middle}}: The same curves from the left panel are now shown as a function of 
    the photon arrival time $t_{\rm obs}$ to an observer at $\theta_{\rm obs}=0$ from emission by 
    material moving along different \rev{$\theta = \{0,0.05\}$}. The two vertical lines show $t_{\rm GRB}$ (for $\theta=0$). 
    \textbf{\textit{Right}}: 
    Proper speed evolution in the thick-shell case shown for different external medium density profiles,
    $n_{\rm ext}=n_0(R/R_0)^{-k}$ with $\Gamma_0=10^3$. The $u_3$ of the RS-heated material changes slope as it 
    transitions from having a relativistic (thick-shell) to a non-relativistic (thin-shell) temperature.
    }
    \label{fig:Gamma-evol}
\end{figure*}

The angular structure of  the relativistic outflows in GRBs is challenging to constrain. 
An important step in this direction came with the discovery of the coincident detection of gravitational waves 
and short prompt GRB emission from the binary neutron star merger in GW170817 \citep{Abbott+17-GW170817-Ligo-Detection}, 
followed by a peculiar afterglow \citep[e.g.,][]{Margutti+18,Troja+19}. \rev{Hints of jet angular structure 
were also found in other GRBs \citep[e.g.][]{Racusin+08}.} 
An angular structure naturally forms as the jet bores its way through the confining medium near its launching site,
i.e. the stellar envelope in long-soft GRBs 
and the binary merger ejecta in short-hard GRBs, as has been demonstrated with numerical simulations \citep[e.g.,][]{Gottlieb+20,Gottlieb+21,Nathanail+21}. 
More importantly, the jet angular structure is sensitive to the amount of mixing that occurs during jet breakout, between 
the inner and outer cocoon (shocked jet material and confining medium).
The amount of mixing is, in turn, strongly affected by the jet's initial magnetization, 
which remains 
poorly constrained 
and its knowledge is of paramount importance.

Afterglow emission arising from the FS 
for different jet angular structures have been calculated in several works 
\citep[e.g.][]{Rossi+02,Rossi+04,Granot-Kumar-03,Kumar-Granot-03,Granot+05,Granot-05,Gill-Granot-18b,Gill-Granot-20,Beniamini+20,Beniamini+22} 
with varying degree of sophistication. The reverse shock emission from such jets with angular structure was considered 
only in few works \citep[][]{Yan+07,Lamb-Kobayashi-19}. Here we systematically consider the reverse shock emission from 
jets with angular structure, along with the system's global RS-FS dynamics,
which becomes 
important in the thick-shell regime (see \S\ref{sec:FS-RS-dynamics}). 

The multi-waveband afterglow observations of the exceptionally bright \grb~present a golden opportunity to constrain the jet 
angular structure and its initial magnetization. We start by summarizing the analytic scalings for thin and thick shells in 
\S\ref{sec:FS-RS-dynamics} for later comparison. In \S\ref{sec:dynamics} we calculate the FS and RS dynamical evolution in an adiabatic spherical 
blast wave using our semi-analytical model. We extend this model to angular structured jets and obtain the afterglow emission 
from both shocks in \S\ref{sec:model-fit}. By \rev{comparing the emission} to time-resolved spectra and lightcurve of \grb~we demonstrate that 
its afterglow is produced by a jet with a shallow angular structure. The implications of our results are discussed in \S\ref{sec:discussion}.

\section{Forward-Reverse Shock Dynamics}\label{sec:FS-RS-dynamics}
The self-similar dynamical evolution and radial structure of a relativistic spherical blast wave was first 
calculated by \citet[][BM76 hereafter]{Blandford-McKee-76}. It was used to explain the long-lasting and broad band afterglow 
emission of GRBs in several seminal works \citep[e.g.][]{Rees-Meszaros-92,Meszaros-Rees-93,Sari-Piran-95}. 
In this model, a cold baryonic spherical shell of (lab-frame) initial width $\Delta_0$ and with kinetic energy $E_{k,\rm iso}$ 
coasting at an initial bulk-LF $\Gamma_0$ is decelerated by sweeping up the surrounding external medium with number density 
$n_{\rm ext}(R)=n_0(R/R_0)^{-k}$. In the process a double shock structure develops with four different regions (quantities in these 
regions have their respective labels): (1) the unshocked external medium, (2) the shocked external medium, (3) the shocked ejecta, 
and (4) the unshocked ejecta. Regions 2 and 3 are separated by a contact discontinuity (CD) and thus have the same pressure and LF, $\Gamma_2=\Gamma_3$.
A relativistic FS propagates into region 1
with bulk 
LF $\Gamma_{\rm sh}=\sqrt{2}\Gamma_2$, shock heating the swept up external medium and accelerating it to a LF $\Gamma_2$. 
At the same time, a 
RS, which may either be relativistic (thick shell) or not (thin shell), propagates into the relativistic ejecta, shock heating and decelerating it. 

The dynamical evolution of this adiabatically expanding system \citep{Sari-Piran-95,Kobayashi+99} depends on whether the reverse shock becomes 
relativistic before crossing the ejecta, the so called \textit{thick-shell} scenario for which $\Gamma_4=\Gamma_0>\Gamma_{\rm cr}$, 
or remains Newtonian (or becomes at least mildly relativistic) as in the \textit{thin-shell} case for which $\Gamma_0<\Gamma_{\rm cr}$. 
The two regimes are separated by the critical LF \citep[e.g.,][]{Kobayashi-Zhang-03,Granot-Ramirez-Ruiz-12,Granot-12a}, 
$\Gamma_{\rm cr}=(\ell/\Delta_0)^{(3-k)/2(4-k)}$, where $\ell\equiv(E_{k,\rm iso}/Ac^2)^{1/(3-k)}$ is the Sedov length 
(ignoring order unity factors), $A=m_pn_0R_0^k$, and $m_p$ is the proton mass. The initial width of the shell can be inferred 
from $\Delta_0=(1+z)^{-1}ct_{\rm GRB}$ for a prompt GRB of apparent duration $t_{\rm GRB}$ at a redshift $z$.

After the RS becomes relativistic in the thick-shell case, 
it starts to considerably reduce the ejecta's bulk LF ($\Gamma_3\ll\Gamma_4$). 
During this time, the LF of the shocked material (contact discontinuity) is independent of $\Gamma_0$ and 
obeys the scaling $\Gamma_3=\Gamma_2\sim(L/Ac^3)^{1/4}R^{-(2-k)/4}$
\citep{Sari-97} where $L=E_{k,\rm iso}c/\Delta_0$ is the source's (isotropic equivalent) kinetic power. 
After that, only a small fraction of the initial kinetic energy of the ejecta is extracted at a radius where 
the mass of the swept up external medium becomes comparable to $M_0/\Gamma_0$, with $M_0=E_{k,\rm iso}/\Gamma_0c^2$ 
being the baryon load of the ejecta. Most of this energy is extracted at the shock crossing radius, $R_\Delta\sim R_{\rm cr}\sim\Delta_0\Gamma_{\rm cr}^2$, where 
the shocked ejecta and shocked external medium have comparable energies.
Beyond $R_{\rm cr}$,
most of the initial kinetic energy of the ejecta resides in the shocked external medium while the fraction in the 
shocked ejecta rapidly declines. The evolution of the blast wave (FS) is then described by the self-similar BM76 solution 
such that $\Gamma_2\sim\Gamma_{\rm cr}(R/R_{\rm cr})^{-(3-k)/2}\sim(E_{k,\rm iso}/Ac^2)^{1/2}R^{-(3-k)/2}$. So long the shocked ejecta is relativistically 
hot (with adiabatic index $\hat\gamma=4/3$), its evolution is approximately consistent with the BM76 self-similar radial profile 
\citep{Sari-Piran-99a,Sari-Piran-99b,Kobayashi+99,Kobayashi-Sari-00}, with $\Gamma_3\sim\Gamma_{\rm cr}(R/R_{\rm cr})^{-(7-2k)/2}$. 
When the temperature of the shocked ejecta becomes non-relativistic, the BM76 solution no longer applies to it.

In the thin-shell case, the reverse shock never becomes relativistic. It is initially Newtonian and for a reasonable spread 
in the initial LF, $\Delta\Gamma_0\sim\Gamma_0$, it starts spreading radially at $R_s\sim\Delta_0\Gamma_0^2$  with 
$\Delta\sim\max(\Delta_0,R/\Gamma_0^2)$, and
becomes mildly 
relativistic near the crossing radius, $R_\Gamma\sim(E_{k,\rm iso}/Ac^2\Gamma_0^2)^{1/(3-k)}\sim\ell\,\Gamma_0^{-2/(3-k)}$. 
Therefore, at $R_\Gamma$ we have
$\Gamma_3=\Gamma_2\approx\Gamma_0$, the swept up mass 
is $\sim M_0/\Gamma_0$, and
most of the kinetic energy is transferred to the 
shocked external medium.
Beyond $R_\Gamma$ the blast wave (FS) commences the BM76 self-similar evolution. 
Since the reverse shock never becomes relativistic, the shocked ejecta remains at a sub-relativistic temperature 
(with $\hat\gamma=5/3$). As a result, its evolution cannot be described by the BM76 solution after shock crossing. 
Assuming a power law with $\Gamma_3(R)\propto R^{-g}$, \citet{Kobayashi-Sari-00} argued that $3/2\leq g\leq7/2$ 
(or $\frac{3-k}{2}\leq g\leq\frac{7-2k}{2}$ for a general $k$) and showed that $g\sim2$ is consistent with 
one-dimensional hydrodynamical numerical simulations.

\section{Adiabatic Spherical Blast Waves}\label{sec:dynamics}

Many works have considered the dynamical evolution of an impulsive relativistic blast wave undergoing adiabatic and/or radiative expansion 
\citep{Blandford-McKee-76,Katz-Piran-97,Chiang-Dermer-99,Piran-99,Huang+99,Beloborodov-Uhm-06,Peer-12} mostly 
in the limit of a homogeneous shell where 
its radial structure is ignored. Most of these works have focused on the thin-shell scenario where the reverse shock contribution 
to the system dynamics is relatively modest. However, including the influence of the reverse shock on the system dynamics is crucial in the
thick shell case, which is most relevant for many long GRBs, and makes the model more realistic in the thin shell case.
Here we adopt the self-consistent model of 
\citet{Nava+13} (also see \citealt{Zhang+22}) that treats the entire dynamical evolution of an adiabatic/radiative blast wave and 
allows for an arbitrary density profile of the external medium.  
Below we only present its most salient features.

Consider a relativistic spherical shell of initial (lab-frame) thickness $\Delta_0$, bulk LF $\Gamma_0$, and mass loading $M_0$ expanding 
into an external medium with number density $n_{\rm ext}(R)$. The (lab-frame) total energy of the system is
$E_{\rm tot} = \Gamma_0M_{0,4}c^2 + \Gamma M_{0,3}c^2 + \Gamma mc^2 + \Gamma_{\rm eff,3}E_{\rm int,3}' + \Gamma_{\rm eff,2}E_{\rm int,2}'$
where $\Gamma_4=\Gamma_0$ and $M_0 = M_{0,3}+M_{0,4}$. During reverse-shock crossing it is assumed that both regions (2) and (3) are in pressure 
balance ($P_2'=P_3'$) and move with the same bulk LF ($\Gamma_3=\Gamma_2\equiv\Gamma$). 
The total mass swept up by the blast wave as it propagates a radial distance $R$ in an external medium with 
density $n_{\rm ext}(R)\propto R^{-k}$ is $m(R)=[4\pi m_p/(3-k)]R^3 n_{\rm ext}(R)$ 
for $k<3$. The comoving internal energy of the shocked material is 
given by $E_{\rm int}'$ and $\Gamma_{\rm eff}=(\hat\gamma\Gamma^2-\hat\gamma+1)/\Gamma$ is the appropriate LF to accurately 
describe the Lorentz transformation of the internal energy \citep{Peer-12}. Here $\hat\gamma = (4\Gamma_{ud} + 1)/3\Gamma_{ud} = 4/3(5/3)$ 
is the adiabatic index for a fluid with relativistic (non-relativistic) temperature and $\Gamma_{ud} = \Gamma_u\Gamma_d(1-\beta_u\beta_d)$ 
is the relative LF between the upstream and downstream material that have LFs $\Gamma_{\{u,d\}} = (1-\beta_{\{u,d\}}^2)^{-1/2}$. 

As the shell propagates an infinitesimal radial distance $dR$, its total energy changes by an amount 
$dE_{\rm tot} = dmc^2 + \Gamma_{\rm eff,2}dE_{\rm rad}'$ 
that is a sum of the rest mass energy of the matter swept up over that interval, $dm = 4\pi m_p R^2 n_{\rm ext}dR$, and the energy that is 
radiated away. For an adiabatic blast wave, radiative losses are negligible over the dynamical time, and therefore, the second term 
can be neglected. The change in the internal energy of the shocked material is a sum of three terms, 
$dE_{\rm int}' = dE_{\rm sh}' + dE_{\rm ad}' + dE_{\rm rad}'$. Here $dE_{\rm sh}' = (\Gamma_{ud}-1)dMc^2$ is the random kinetic energy 
of the newly shocked material, where $dM=dm$ for region (2) and $dM=dM_{0,3}$ for region (3). The terms $dE_{\rm ad}'$ and $dE_{\rm rad}'$ 
represent the change in energy due to adiabatic and radiative losses (see Eq.\,(13) of \citet{Nava+13}). By taking the differential of $E_{\rm tot}$ 
and noticing that $dM_{0,4}=-dM_{0,3}$, the rate of change of the bulk LF of the shocked material can be obtained,
\begin{eqnarray}\label{eq:dG_dR}
    \frac{d\Gamma}{dR} &=& -\frac{(\Gamma_{\rm eff,2}+1)(\Gamma-1)\frac{dm}{dR}c^2+\Gamma_{\rm eff,2}\frac{dE_{\rm ad,2}'}{dR}}
    {(M_{0,3}+m)c^2+E_{\rm int,2}'\frac{d\Gamma_{\rm eff,2}}{d\Gamma}+E_{\rm int,3}'\frac{d\Gamma_{\rm eff,3}}{d\Gamma}} \\
    &&-\frac{(\Gamma-\Gamma_0-\Gamma_{\rm eff,3}+\Gamma_{\rm eff,3}\Gamma_{43})\frac{dM_{0,3}}{dR}c^2+\Gamma_{\rm eff,3}\frac{dE_{\rm ad,3}'}{dR}}
    {(M_{0,3}+m)c^2+E_{\rm int,2}'\frac{d\Gamma_{\rm eff,2}}{d\Gamma}+E_{\rm int,3}'\frac{d\Gamma_{\rm eff,3}}{d\Gamma}}\,. \nonumber
\end{eqnarray}
This equation needs to be supplemented by $dM_{0,3}/dR$, the rate at which ejecta mass is entering the reverse shock. The differential 
number of baryons shocked by the RS as it moves a (lab-frame) radial distance $d\Delta$ into the ejecta is $dN_4=4\pi R^2n_4'\Gamma_4d\Delta$. Here 
we explicitly assume that the width of the shocked region is much smaller than the radial distance of the blast wave, and therefore, 
both shocks are approximately at the same radius, such that $R_{\rm RS}\sim R_{\rm FS}=R$. Conservation of baryon number dictates that 
$d\Delta = (\beta_4-\beta_3)(1-\Gamma_4n_4'/\Gamma_3n_3')^{-1}dR$ \citep{Sari-Piran-95}. Putting this all together yields the mass injection 
rate into region (3), $dM_{0,3}/dR = 4\pi R^2\rho_4'\Gamma_0(\beta_0-\beta)\left(1-\Gamma_0\rho_4'/\Gamma\rho_3'\right)^{-1}$
where $\rho_4'=M_0/4\pi R^2\Delta\Gamma_0$ is the proper mass density of the ejecta shell that has lab-frame width 
$\Delta = \max[\Delta_0,R/\Gamma_0^2]$.

We solve the system of equations describing $d\Gamma/dR$ and $dM_{0,3}/dR$ using a fourth-order Runge-Kutta-Fehlberg (RK45) 
routine with adaptive step size. Until RS crossing $\Gamma_3(R)=\Gamma_2(R)=\Gamma(R)$, but after RS crossing the evolution of the 
shocked ejecta is prescribed to either have a BM76 profile (thick-shell), with $u_3(R)=\Gamma_3(R)\beta_3(R)=u_\Delta(R/R_\Delta)^{-(7-2k)/2}$, 
or a general power-law (thin-shell), with $u_3(R)=u_\Delta(R/R_\Delta)^{-g}$. The shock crossing radius ($R_\Delta$) is determined 
by integrating the equation for $dM_{0,3}/dR$ and by implementing the condition $M_{0,3}(R_\Delta)=M_0$, and the corresponding $u_\Delta = 
u(R_\Delta)$ is obtained from the solution of Eq.\,(\ref{eq:dG_dR}).

The BM76 profile is strictly valid for a relativistically hot gas that has adiabatic index $\hat\gamma=4/3$. As the flow expands, the 
comoving pressure and number density decline as $P_3'\propto R^{-2(13-2k)/3}$ and $n_3'\propto R^{-(13-2k)/2}$. The sound speed in the 
comoving frame, $c_s'/c = (\hat\gamma P'/w')^{1/2}$, depends on the ratio $w'/P'=(\rho'/P')c^2 + \hat\gamma/(\hat\gamma-1)$ 
where $w'$ is the enthalpy density. In the thick-shell case, the shocked gas is initially relativistically hot, which means that 
$(\rho_3'/P_3')c^2\ll\hat\gamma/(\hat\gamma-1)$ and therefore, $c_{s,3}'/c=\sqrt{\hat\gamma_3-1}=1/\sqrt{3}$. The ratio 
$\rho_3'/P_3'\propto R^{(13-2k)/6}$ grows with radius, and as a result, a relativistically hot fluid will become non-relativistic. 
Therefore, we switch to the thin-shell evolution when $(\rho_3'/P_3')c^2>\hat\gamma_3/(\hat\gamma_3-1)=4$. 

In Fig.\,\ref{fig:Gamma-evol}, we show the 
proper speed $u=\Gamma\beta$ evolution of the shocked material downstream of both 
the RS and FS for a thin and thick shell (\textit{left} \& \textit{middle} panels) and for different external medium density 
profiles in the thick shell case (\textit{right} panel). 
We compare the asymptotic slopes of $u(R)$ in different regimes with analytical results and find excellent agreement. The model 
evolves the bulk LFs of the shocked material downstream of both the shocks until the crossing radius $R_\Delta$ and 
only for the material behind the FS thereafter. For $R>R_\Delta$, the $u_3(R)$ profiles are manually prescribed. 
In the thick-shell case, $u_3(R)$ shows a change in slope as the shocked material transitions from having a relativistic 
(thick-shell) to a non-relativistic (thin-shell) temperature.


\begin{table*}
    \centering
    \begin{tabular}{c|c|c|c|c|c|c|c|c|c|c|c|c|c|c|c}
    \hline
        $\theta_{\rm obs}$ & \rev{$E_{k,\rm {iso},c}$} & \rev{$\Gamma_c$} & $n_0(R_0)$ & $k$ & $A_*$ & $p_{\rm FS}$ & $\epsilon_{e,-2}^{\rm FS}$ & $\epsilon_{B,-4}^{\rm FS}$ & $\xi_{e,\rm FS}$ & 
        $p_{\rm RS}$ & $\epsilon_{e,-2}^{\rm RS}$ & $\epsilon_{B,-4}^{\rm RS}$ & $\xi_{e,\rm RS}$ & $g$ & $t_{\rm GRB}$ \\
        \hline
         $0.02$\,rad & $2\times10^{55}\,$erg & 300 & $0.1\,{\rm cm}^{-3}$ & $2$ & $0.33$ & $2.4$ & $1.0$ & $1.0$ & $0.01$ & 
         $2.03$ & $8$ & $10$ & $0.01$ & $1.3$ & $500\,$s \\
         \hline
    \end{tabular}
    \caption{Model parameters. The external density is normalized at $R_0=10^{18}$\,cm 
    \rev{($A_*=n_0R_0^2/3\times10^{35}\,{\rm cm}^{-1}$); $(a,\,b,\,\theta_{c,\epsilon},\,\theta_{c,\Gamma})=(0.8,\,0.3,\,0.021,\,0.016)$}.
    }
    \label{tab:params}
\end{table*}

\section{The Afterglow of GRB 221009A}
\label{sec:model-fit}
We consider an ultra-relativistic outflow with a power-law angular structure to describe the 
afterglow of \grb. \rev{This type of structure has been obtained in numerical simulations of long (short) GRB jets penetrating out of the 
progenitor star (merger ejecta) \citep[e.g.][]{Gottlieb+21}, and was successful in explaining the peculiar afterglow of 
GW\,170817/GRB\,170817A \citep[e.g.][]{Gill-Granot-18b}.}  

\begin{figure}
    \centering
\includegraphics[trim={0cm 0cm 0cm 0cm},clip,scale=0.6,width=0.48\textwidth]{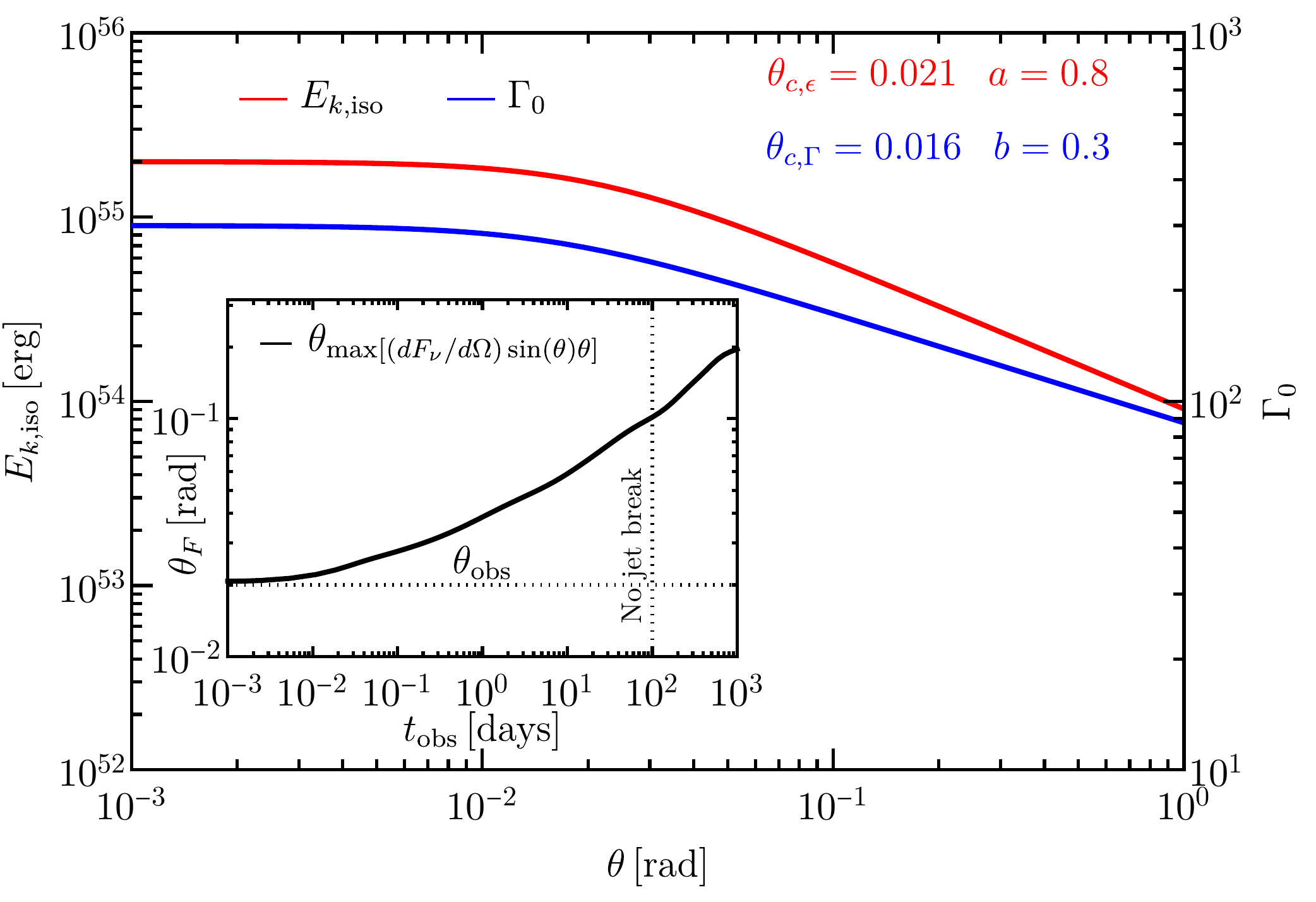}
\vspace{-0.6cm}
    \caption{Angular structure of the outflow, showing the isotropic-equivalent (total) kinetic energy 
    and initial bulk LF profiles as a function of polar angle $\theta$ measured from the jet symmetry 
    axis. The inset shows the temporal evolution of the polar angle $\theta_F$ that dominates the X-ray flux, 
    \rev{where the latter has shown no steepening due to a jet break at $t_{\rm obs}<100\,$days.}    
    }
    \label{fig:jet-structure}
\end{figure}

Both the energy per unit solid angle, $\epsilon(\theta)=dE(\theta)/d\Omega = E_{k,\rm iso}(\theta)/4\pi$ where $d\Omega$ is the 
element of the solid angle, and the initial bulk LF (minus one) are described by a power law at polar angle $\theta$ 
(measured from the jet symmetry axis) larger than some core angle $\theta_c$ 
\citep[e.g.,][]{Granot-Kumar-03}
\begin{equation}
    \left\{\frac{\epsilon}{\epsilon_c}, \frac{\Gamma_0(\theta)-1}{\Gamma_c-1}\right\} = \left[1+\fracb{\theta}{\theta_{c,\{\epsilon,\Gamma\}}}^2\right]^{-\{a,b\}/2}\,.
\end{equation}
The core angle and power-law indices for the energy and bulk LF are allowed to be different. 
Here we apply our model of a spherical blast wave to an angular structured flow by 
assuming that every angle $\theta$ on the flow's surface evolves independently as if it were part of a spherical 
flow with kinetic energy $E_{k,\rm iso}(\theta)$ and initial LF $\Gamma_0(\theta)$. Its dynamical evolution 
is obtained by solving the equations used for the spherical flow but now on a grid of polar angles $\theta$. 
For simplicity and computational convenience we ignore lateral spreading, which becomes important as the outflow 
approaches the non-relativistic Sedov-Taylor phase.
In Fig.\,\ref{fig:jet-structure} we show the angular structure of the outflow that was adopted in order to describe the 
afterglow of \grb. 

To calculate the afterglow emission we adopt the treatment in \citet{Gill-Granot-18b} and assume for simplicity that 
it arises from an infinitely thin shell.
The two shocks accelerate a fraction $\xi_e$ of the total swept up electrons into a power-law energy distribution 
with $dn_e'/d\gamma_e\propto\gamma_e^{-p}$ (for $\gamma_e>\gamma_m$), where $n_e'$ is the number density of the synchrotron emitting electrons 
and $\gamma_e$ is their LF. These electrons receive a fraction $\epsilon_e$ of the total internal energy of the shocked 
gas, whereas a fraction $\epsilon_B$ of the same goes into generating the small scale magnetic field that leads to 
synchrotron cooling. At a given 
observed time $t_{\rm obs}$, the emission is obtained by integrating over the equal-arrival-time surface (EATS). 
We obtain smoother and more realistic spectral breaks in the comoving spectrum by adopting the smoothing prescription 
introduced in \citet{Granot-Sari-02} for $k=\{0,2\}$ and later generalized to $0\leq k\leq 2$ in \citet{Leventis+12}. 

\rev{We find that the viewing angle of the observer is comparable to the core angle of the energy profile, 
$\theta_{\rm obs}\sim\theta_{c,\epsilon}$, which is needed in order to explain the high fluence of the $\gamma$-ray emission 
and the multi-waveband lightcurve. In contrast to some works \citep[e.g.][]{OConner+23}, we find $k\approx2$ 
to avoid overproducing the sub-mm flux by the FS emission. The kinetic energy of the bipolar outflow in our model 
is $E_k\propto\theta_F^{2-a}\propto t_{\rm obs}^{(2-a)(3-k)\over(8-2k-a)}\simeq4.2\times10^{52}(t_{\rm obs}/100\,{\rm days})^{0.375}\,$erg 
for no jet break at $t_{\rm obs}<100\,$days \citep[see][for scalings]{Beniamini+22}, with $E_{k,\max}\simeq7\times10^{53}\,$erg 
for a maximum jet angular size of $\theta_{\max}=1\,$rad.} All the model parameters adopted for the FS and RS regions are 
shown in Table\,\ref{tab:params}. \rev{These illustrate one possible approximate solution, while significant degeneracy remains.}

\begin{figure*}\label{fig:spectrum}
    \centering
    \includegraphics[width=0.645\textwidth]{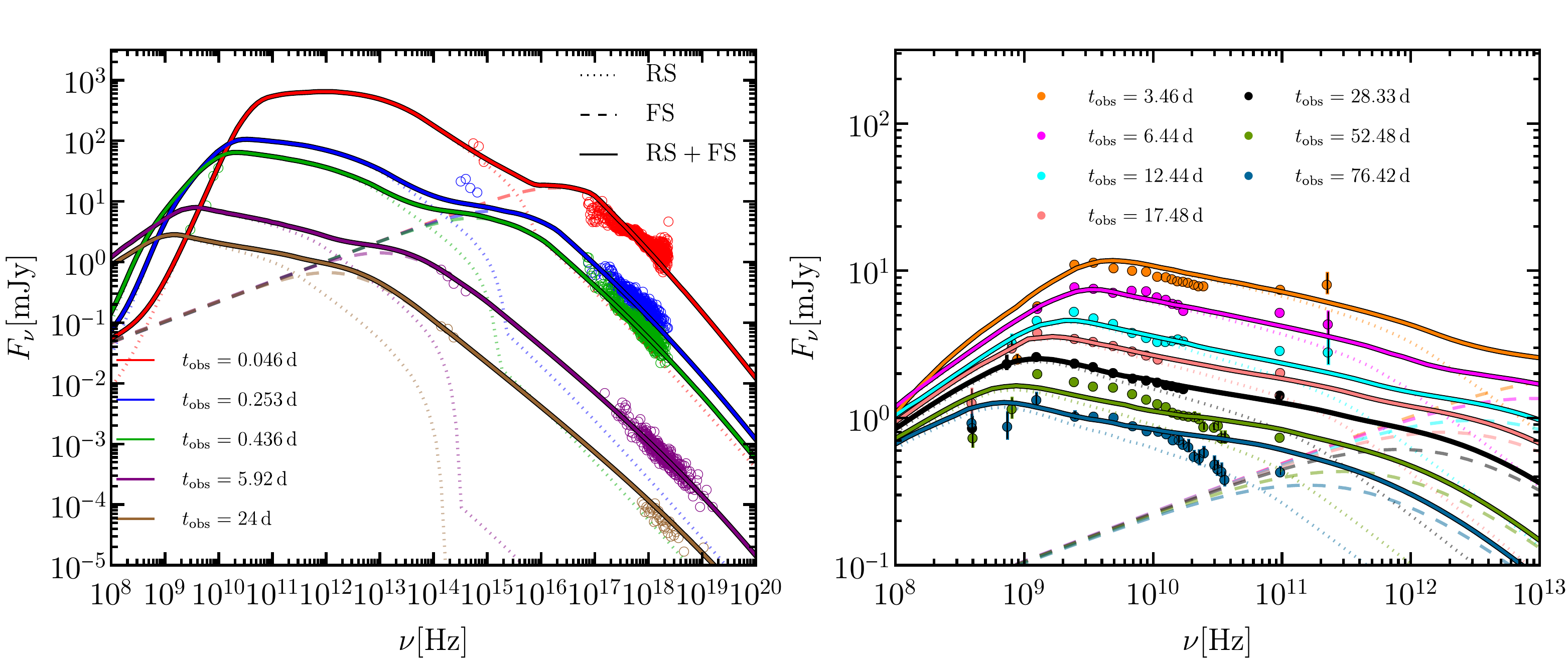}\quad
    \includegraphics[width=0.315\textwidth]{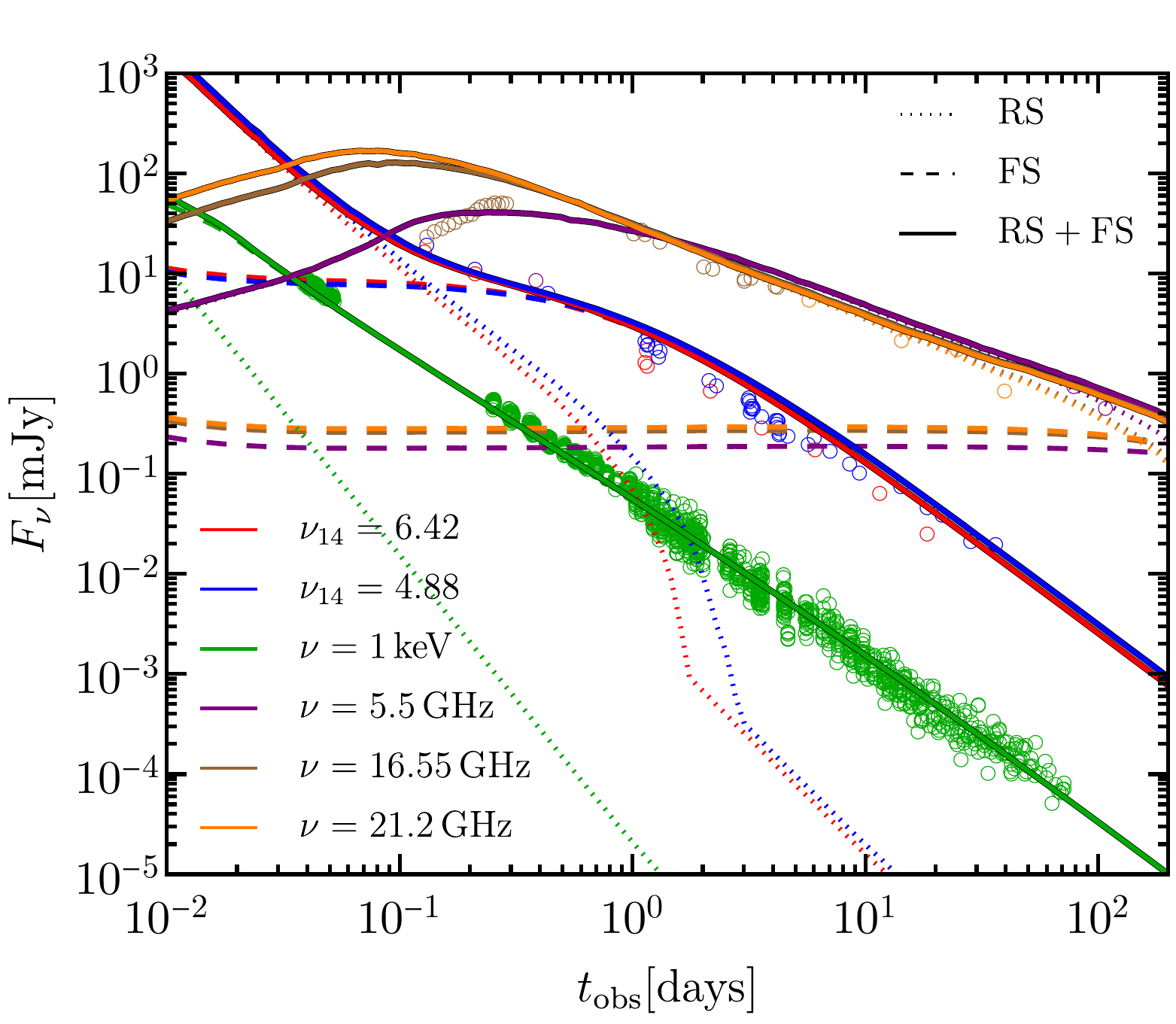}
    \vspace{-0.3cm}
    \caption{\textbf{\textit{Left}}: \rev{Comparison of} model spectra with the multi-wavelength afterglow of \grb~from radio to X-rays 
    at different observed times $t_{\rm obs}$ (measured from the GRB trigger time). 
    \textbf{\textit{Middle}}: Model \rev{comparison to} the radio data from \citet{Laskar+23}. The total emission is shown with solid 
    lines while the reverse shock (RS) and forward shock (FS) components are shown with dotted and dashed lines, 
    respectively.
    \textbf{\textit{Right}}: Model \rev{comparison} to the multi-waveband lightcurve. 
    The X-ray data is from \textit{Swift}/XRT, and the optical and radio data are from \citet{OConner+23,Bright+23}.
    }
\end{figure*}

The left panel of Fig.\,\ref{fig:spectrum} \rev{compares} the model \rev{spectra with} multi-wavelength afterglow 
\rev{observations} of \grb~at different times and the right panel shows the corresponding \rev{comparison} to the lightcurve. 
The X-ray afterglow is completely dominated by the FS emission at all times. The optical afterglow is initially RS 
dominated, at $t_{\rm obs}\lesssim0.1\,$days, and then becomes increasingly FS dominated. The RS emission also 
becomes highly suppressed at higher frequencies as the cut-off frequency ($\nu_{\rm cut}$) passes through the optical 
band towards even lower frequencies over time. The emission is not completely suppressed at $\nu>\nu_{\rm cut}$, 
as one would expect from analytic models that only account for emission along the line-of-sight (LOS). Due to integration 
over the EATS, high-latitude emission coming from angles $\vert\theta-\theta_{\rm obs}\vert>1/\Gamma(\theta_{\rm obs})$ and 
which was emitted at earlier lab-frame times, when $\nu_{\rm cut}>\nu$, makes a dominant contribution to the total flux over that emitted along the LOS. 
This effect will be described 
in a future publication (Gill \& Granot, in preparation). 

The middle panel of Fig.\,\ref{fig:spectrum} \rev{compares the model spectra} to the radio afterglow data from \citet{Laskar+23}. 
The RS emission clearly dominates the radio flux at all times, with significant FS contribution to the sub-millimeter 
flux coming in at late times. The RS radio emission is also self-absorbed below the self-absorption frequency, which 
decreases with time. 

Our model is able to describe the multi-waveband lightcurve reasonably well. \rev{In fact,} the FS dominated X-ray lightcurve \rev{compares 
remarkably well with } the data over a wide range of timescales and captures the shallow break around $t_{\rm br}\sim0.8\,$days. 
This break is caused by the shallow break in the energy profile at $\theta\sim0.04\,$rad, emission from which angle 
dominates the flux at $t_{\rm obs}=t_{\rm br}$ (see inset in Fig\,\ref{fig:jet-structure}). A similar conclusion was reached in 
\citet{OConner+23}. At $t_{\rm obs}\lesssim0.3\,$days, $\nu_{\rm opt}<\nu_m$ and therefore the FS lightcurve is expected to be flat when 
$k=2$. This means that the slowly decaying optical flux at early times must come from a diminishing RS contribution 
The radio data, which is completely 
dominated by emission from the RS, is explained well by our model at $t_{\rm obs}\gtrsim1\,$day. At earlier times, 
it over predicts the radio flux by a factors of a few when compared with the radio data from \citet{Bright+23}. 
The origin of this discrepancy may lie in the rather coarse exploration of parameter space degeneracies conducted here. 
This can be addressed more efficiently using a simpler and computationally less intense parameterized model \citep[e.g.,][]{Lamb-Kobayashi-19}.

\vspace{-0.1cm}
\section{Conclusions}
\label{sec:discussion}
We have explored a potential solution of emission from a jet with a shallow angular profile, 
both in energy ($a\approx0.8$) and initial bulk LF ($b\approx0.3$), to explain the multi-waveband 
afterglow of the exceptionally bright \grb. In this model, the radio emission 
is attributed to that arising from the RS heated ejecta while the optical and X-ray flux is coming from 
the FS heated external medium. 

We find that \revv{the data appear to favor that only a small ($\xi_e\approx10^{-2}$) fraction of shock-heated electrons are accelerated to a power-law energy distribution in both shocks.}
Similar, but somewhat larger, fractions 
have been obtained in other works \citep[e.g.][]{Salafia+22} and such a result is consistent with particle-in-cell 
simulations of magnetized collisionless shocks \citep{Sironi-Spitkovsky-11}.

This is the first GRB in which clear hints for a shallow \rev{($a\lesssim2$)} structured outflow have been found. 
One of the key observations is the presence of a shallower achromatic break 
in the lightcurve ($F_\nu\propto t_{\rm obs}^\alpha$) with $\Delta\alpha\sim0.14$ at $t_{\rm obs}\sim0.8$\,days 
as compared to $\Delta\alpha=\frac{3-k}{4-k}=0.75\,(0.5)$ for $k=0\,(2)$, the jet-break expected 
in a non-spreading sharp-edged jet. \citet{OConner+23} pointed out several other bright jet-break-less 
GRBs that may be explained using the model explored here. If true, it may offer strong support 
for weakly magnetized jets in at least some long-soft GRBs \citep{Bromberg-Tchekhovskoy-16} 
as they break out from the stellar envelope, leading to more mixing and a shallow energy angular 
profile ($a\lesssim2$).

\vspace{-0.4cm}
\section*{Acknowledgements}

R.G. acknowledges financial support from the UNAM-DGAPA-PAPIIT IA105823 grant, Mexico.
J.G. acknowledges financial support by the ISF-NSFC joint research program (grant no. 3296/19).

\vspace{-0.4cm}
\section*{Data Availability}

The model data can be shared upon reasonable request to the authors.


\vspace{-0.3cm}
\bibliographystyle{mnras}
\bibliography{refs} 

\begin{thebibliography}{}
\makeatletter
\relax
\def\mn@urlcharsother{\let\do\@makeother \do\$\do\&\do\#\do\^\do\_\do\%\do\~}
\def\mn@doi{\begingroup\mn@urlcharsother \@ifnextchar [ {\mn@doi@}
  {\mn@doi@[]}}
\def\mn@doi@[#1]#2{\def\@tempa{#1}\ifx\@tempa\@empty \href
  {http://dx.doi.org/#2} {doi:#2}\else \href {http://dx.doi.org/#2} {#1}\fi
  \endgroup}
\def\mn@eprint#1#2{\mn@eprint@#1:#2::\@nil}
\def\mn@eprint@arXiv#1{\href {http://arxiv.org/abs/#1} {{\tt arXiv:#1}}}
\def\mn@eprint@dblp#1{\href {http://dblp.uni-trier.de/rec/bibtex/#1.xml}
  {dblp:#1}}
\def\mn@eprint@#1:#2:#3:#4\@nil{\def\@tempa {#1}\def\@tempb {#2}\def\@tempc
  {#3}\ifx \@tempc \@empty \let \@tempc \@tempb \let \@tempb \@tempa \fi \ifx
  \@tempb \@empty \def\@tempb {arXiv}\fi \@ifundefined
  {mn@eprint@\@tempb}{\@tempb:\@tempc}{\expandafter \expandafter \csname
  mn@eprint@\@tempb\endcsname \expandafter{\@tempc}}}

\bibitem[\protect\citeauthoryear{{Abbott} et~al.,}{{Abbott}
  et~al.}{2017}]{Abbott+17-GW170817-Ligo-Detection}
{Abbott} B.~P.,  et~al., 2017, \mn@doi [Physical Review Letters]
  {10.1103/PhysRevLett.119.161101}, \href
  {http://adsabs.harvard.edu/abs/2017PhRvL.119p1101A} {119, 161101}

\bibitem[\protect\citeauthoryear{{An} et~al.,}{{An} et~al.}{2023}]{An+23}
{An} Z.-H.,  et~al., 2023, \mn@doi [arXiv e-prints]
  {10.48550/arXiv.2303.01203}, \href
  {https://ui.adsabs.harvard.edu/abs/2023arXiv230301203A} {p. arXiv:2303.01203}

\bibitem[\protect\citeauthoryear{{Beloborodov} \& {Uhm}}{{Beloborodov} \&
  {Uhm}}{2006}]{Beloborodov-Uhm-06}
{Beloborodov} A.~M.,  {Uhm} Z.~L.,  2006, \mn@doi [\apjl] {10.1086/508807},
  \href {https://ui.adsabs.harvard.edu/abs/2006ApJ...651L...1B} {651, L1}

\bibitem[\protect\citeauthoryear{{Beniamini}, {Granot}  \& {Gill}}{{Beniamini}
  et~al.}{2020}]{Beniamini+20}
{Beniamini} P.,  {Granot} J.,   {Gill} R.,  2020, \mn@doi [\mnras]
  {10.1093/mnras/staa538}, \href
  {https://ui.adsabs.harvard.edu/abs/2020MNRAS.493.3521B} {493, 3521}

\bibitem[\protect\citeauthoryear{{Beniamini}, {Gill}  \& {Granot}}{{Beniamini}
  et~al.}{2022}]{Beniamini+22}
{Beniamini} P.,  {Gill} R.,   {Granot} J.,  2022, \mn@doi [\mnras]
  {10.1093/mnras/stac1821}, \href
  {https://ui.adsabs.harvard.edu/abs/2022MNRAS.515..555B} {515, 555}

\bibitem[\protect\citeauthoryear{{Blandford} \& {McKee}}{{Blandford} \&
  {McKee}}{1976}]{Blandford-McKee-76}
{Blandford} R.~D.,  {McKee} C.~F.,  1976, \mn@doi [Physics of Fluids]
  {10.1063/1.861619}, \href {http://adsabs.harvard.edu/abs/1976PhFl...19.1130B}
  {19, 1130}

\bibitem[\protect\citeauthoryear{{Bright} et~al.,}{{Bright}
  et~al.}{2023}]{Bright+23}
{Bright} J.~S.,  et~al., 2023, \mn@doi [arXiv e-prints]
  {10.48550/arXiv.2303.13583}, \href
  {https://ui.adsabs.harvard.edu/abs/2023arXiv230313583B} {p. arXiv:2303.13583}

\bibitem[\protect\citeauthoryear{{Bromberg} \& {Tchekhovskoy}}{{Bromberg} \&
  {Tchekhovskoy}}{2016}]{Bromberg-Tchekhovskoy-16}
{Bromberg} O.,  {Tchekhovskoy} A.,  2016, \mn@doi [\mnras]
  {10.1093/mnras/stv2591}, \href
  {https://ui.adsabs.harvard.edu/abs/2016MNRAS.456.1739B} {456, 1739}

\bibitem[\protect\citeauthoryear{{Burns} et~al.,}{{Burns}
  et~al.}{2023}]{Burns+23}
{Burns} E.,  et~al., 2023, \mn@doi [\apjl] {10.3847/2041-8213/acc39c}, \href
  {https://ui.adsabs.harvard.edu/abs/2023ApJ...946L..31B} {946, L31}

\bibitem[\protect\citeauthoryear{{Chiang} \& {Dermer}}{{Chiang} \&
  {Dermer}}{1999}]{Chiang-Dermer-99}
{Chiang} J.,  {Dermer} C.~D.,  1999, \mn@doi [\apj] {10.1086/306789}, \href
  {https://ui.adsabs.harvard.edu/abs/1999ApJ...512..699C} {512, 699}

\bibitem[\protect\citeauthoryear{{Frederiks} et~al.,}{{Frederiks}
  et~al.}{2023}]{Frederiks+23}
{Frederiks} D.,  et~al., 2023, \mn@doi [arXiv e-prints]
  {10.48550/arXiv.2302.13383}, \href
  {https://ui.adsabs.harvard.edu/abs/2023arXiv230213383F} {p. arXiv:2302.13383}

\bibitem[\protect\citeauthoryear{{Fulton} et~al.,}{{Fulton}
  et~al.}{2023}]{Fulton+23}
{Fulton} M.~D.,  et~al., 2023, \mn@doi [arXiv e-prints]
  {10.48550/arXiv.2301.11170}, \href
  {https://ui.adsabs.harvard.edu/abs/2023arXiv230111170F} {p. arXiv:2301.11170}

\bibitem[\protect\citeauthoryear{{Gill} \& {Granot}}{{Gill} \&
  {Granot}}{2018}]{Gill-Granot-18b}
{Gill} R.,  {Granot} J.,  2018, \mn@doi [\mnras] {10.1093/mnras/sty1214}, \href
  {http://adsabs.harvard.edu/abs/2018MNRAS.478.4128G} {478, 4128}

\bibitem[\protect\citeauthoryear{{Gill} \& {Granot}}{{Gill} \&
  {Granot}}{2020}]{Gill-Granot-20}
{Gill} R.,  {Granot} J.,  2020, \mn@doi [\mnras] {10.1093/mnras/stz3340}, \href
  {https://ui.adsabs.harvard.edu/abs/2020MNRAS.491.5815G} {491, 5815}

\bibitem[\protect\citeauthoryear{{Gottlieb}, {Bromberg}, {Singh}  \&
  {Nakar}}{{Gottlieb} et~al.}{2020}]{Gottlieb+20}
{Gottlieb} O.,  {Bromberg} O.,  {Singh} C.~B.,   {Nakar} E.,  2020, \mn@doi
  [\mnras] {10.1093/mnras/staa2567}, \href
  {https://ui.adsabs.harvard.edu/abs/2020MNRAS.498.3320G} {498, 3320}

\bibitem[\protect\citeauthoryear{{Gottlieb}, {Nakar}  \& {Bromberg}}{{Gottlieb}
  et~al.}{2021}]{Gottlieb+21}
{Gottlieb} O.,  {Nakar} E.,   {Bromberg} O.,  2021, \mn@doi [\mnras]
  {10.1093/mnras/staa3501}, \href
  {https://ui.adsabs.harvard.edu/abs/2021MNRAS.500.3511G} {500, 3511}

\bibitem[\protect\citeauthoryear{{Granot}}{{Granot}}{2005}]{Granot-05}
{Granot} J.,  2005, \mn@doi [\apj] {10.1086/432676}, \href
  {http://adsabs.harvard.edu/abs/2005ApJ...631.1022G} {631, 1022}

\bibitem[\protect\citeauthoryear{{Granot}}{{Granot}}{2012}]{Granot-12a}
{Granot} J.,  2012, \mn@doi [\mnras] {10.1111/j.1365-2966.2012.20473.x}, \href
  {https://ui.adsabs.harvard.edu/abs/2012MNRAS.421.2442G} {421, 2442}

\bibitem[\protect\citeauthoryear{{Granot} \& {Kumar}}{{Granot} \&
  {Kumar}}{2003}]{Granot-Kumar-03}
{Granot} J.,  {Kumar} P.,  2003, \mn@doi [\apj] {10.1086/375489}, \href
  {https://ui.adsabs.harvard.edu/abs/2003ApJ...591.1086G} {591, 1086}

\bibitem[\protect\citeauthoryear{{Granot} \& {Ramirez-Ruiz}}{{Granot} \&
  {Ramirez-Ruiz}}{2012}]{Granot-Ramirez-Ruiz-12}
{Granot} J.,  {Ramirez-Ruiz} E.,  2012, {Jets and GRB unification schemes}

\bibitem[\protect\citeauthoryear{{Granot} \& {Sari}}{{Granot} \&
  {Sari}}{2002}]{Granot-Sari-02}
{Granot} J.,  {Sari} R.,  2002, \mn@doi [\apj] {10.1086/338966}, \href
  {http://adsabs.harvard.edu/abs/2002ApJ...568..820G} {568, 820}

\bibitem[\protect\citeauthoryear{{Granot}, {Ramirez-Ruiz}  \& {Perna}}{{Granot}
  et~al.}{2005}]{Granot+05}
{Granot} J.,  {Ramirez-Ruiz} E.,   {Perna} R.,  2005, \mn@doi [\apj]
  {10.1086/431477}, \href
  {https://ui.adsabs.harvard.edu/abs/2005ApJ...630.1003G} {630, 1003}

\bibitem[\protect\citeauthoryear{{Huang}, {Dai}  \& {Lu}}{{Huang}
  et~al.}{1999}]{Huang+99}
{Huang} Y.~F.,  {Dai} Z.~G.,   {Lu} T.,  1999, \mn@doi [\mnras]
  {10.1046/j.1365-8711.1999.02887.x}, \href
  {https://ui.adsabs.harvard.edu/abs/1999MNRAS.309..513H} {309, 513}

\bibitem[\protect\citeauthoryear{{Huang}, {Hu}, {Chen}, {Zha}, {Liu}, {Yao},
  {Cao}  \& {Experiment}}{{Huang} et~al.}{2022}]{Huang+22GCN}
{Huang} Y.,  {Hu} S.,  {Chen} S.,  {Zha} M.,  {Liu} C.,  {Yao} Z.,  {Cao} Z.,
  {Experiment} T.~L.,  2022, GRB Coordinates Network, \href
  {https://ui.adsabs.harvard.edu/abs/2022GCN.32677....1H} {32677, 1}

\bibitem[\protect\citeauthoryear{{Kann} et~al.,}{{Kann} et~al.}{2023}]{Kann+23}
{Kann} D.~A.,  et~al., 2023, \mn@doi [arXiv e-prints]
  {10.48550/arXiv.2302.06225}, \href
  {https://ui.adsabs.harvard.edu/abs/2023arXiv230206225K} {p. arXiv:2302.06225}

\bibitem[\protect\citeauthoryear{{Katz} \& {Piran}}{{Katz} \&
  {Piran}}{1997}]{Katz-Piran-97}
{Katz} J.~I.,  {Piran} T.,  1997, \mn@doi [\apj] {10.1086/304913}, \href
  {https://ui.adsabs.harvard.edu/abs/1997ApJ...490..772K} {490, 772}

\bibitem[\protect\citeauthoryear{{Kobayashi} \& {Sari}}{{Kobayashi} \&
  {Sari}}{2000}]{Kobayashi-Sari-00}
{Kobayashi} S.,  {Sari} R.,  2000, \mn@doi [\apj] {10.1086/317021}, \href
  {https://ui.adsabs.harvard.edu/abs/2000ApJ...542..819K} {542, 819}

\bibitem[\protect\citeauthoryear{{Kobayashi} \& {Zhang}}{{Kobayashi} \&
  {Zhang}}{2003}]{Kobayashi-Zhang-03}
{Kobayashi} S.,  {Zhang} B.,  2003, \mn@doi [\apjl] {10.1086/367691}, \href
  {https://ui.adsabs.harvard.edu/abs/2003ApJ...582L..75K} {582, L75}

\bibitem[\protect\citeauthoryear{{Kobayashi}, {Piran}  \& {Sari}}{{Kobayashi}
  et~al.}{1999}]{Kobayashi+99}
{Kobayashi} S.,  {Piran} T.,   {Sari} R.,  1999, \mn@doi [\apj]
  {10.1086/306868}, \href
  {https://ui.adsabs.harvard.edu/abs/1999ApJ...513..669K} {513, 669}

\bibitem[\protect\citeauthoryear{{Kumar} \& {Granot}}{{Kumar} \&
  {Granot}}{2003}]{Kumar-Granot-03}
{Kumar} P.,  {Granot} J.,  2003, \mn@doi [\apj] {10.1086/375186}, \href
  {https://ui.adsabs.harvard.edu/abs/2003ApJ...591.1075K} {591, 1075}

\bibitem[\protect\citeauthoryear{{Lamb} \& {Kobayashi}}{{Lamb} \&
  {Kobayashi}}{2019}]{Lamb-Kobayashi-19}
{Lamb} G.~P.,  {Kobayashi} S.,  2019, \mn@doi [\mnras] {10.1093/mnras/stz2252},
  \href {https://ui.adsabs.harvard.edu/abs/2019MNRAS.489.1820L} {489, 1820}

\bibitem[\protect\citeauthoryear{{Laskar} et~al.,}{{Laskar}
  et~al.}{2023}]{Laskar+23}
{Laskar} T.,  et~al., 2023, \mn@doi [\apjl] {10.3847/2041-8213/acbfad}, \href
  {https://ui.adsabs.harvard.edu/abs/2023ApJ...946L..23L} {946, L23}

\bibitem[\protect\citeauthoryear{{Lesage} et~al.,}{{Lesage}
  et~al.}{2023}]{Lesage+23}
{Lesage} S.,  et~al., 2023, \mn@doi [arXiv e-prints]
  {10.48550/arXiv.2303.14172}, \href
  {https://ui.adsabs.harvard.edu/abs/2023arXiv230314172L} {p. arXiv:2303.14172}

\bibitem[\protect\citeauthoryear{{Leventis}, {van Eerten}, {Meliani}  \&
  {Wijers}}{{Leventis} et~al.}{2012}]{Leventis+12}
{Leventis} K.,  {van Eerten} H.,  {Meliani} Z.,   {Wijers} R.,  2012, \mn@doi
  [\mnras] {10.1111/j.1365-2966.2012.21994.x}, \href
  {https://ui.adsabs.harvard.edu/abs/2012MNRAS.427.1329L} {427, 1329}

\bibitem[\protect\citeauthoryear{{Malesani} et~al.,}{{Malesani}
  et~al.}{2023}]{Malesani+23}
{Malesani} D.~B.,  et~al., 2023, \mn@doi [arXiv e-prints]
  {10.48550/arXiv.2302.07891}, \href
  {https://ui.adsabs.harvard.edu/abs/2023arXiv230207891M} {p. arXiv:2302.07891}

\bibitem[\protect\citeauthoryear{{Margutti} et~al.,}{{Margutti}
  et~al.}{2018}]{Margutti+18}
{Margutti} R.,  et~al., 2018, \mn@doi [\apjl] {10.3847/2041-8213/aab2ad}, \href
  {http://adsabs.harvard.edu/abs/2018ApJ...856L..18M} {856, L18}

\bibitem[\protect\citeauthoryear{{Meszaros} \& {Rees}}{{Meszaros} \&
  {Rees}}{1993}]{Meszaros-Rees-93}
{Meszaros} P.,  {Rees} M.~J.,  1993, \mn@doi [\apj] {10.1086/172360}, \href
  {https://ui.adsabs.harvard.edu/abs/1993ApJ...405..278M} {405, 278}

\bibitem[\protect\citeauthoryear{{Nathanail}, {Gill}, {Porth}, {Fromm}  \&
  {Rezzolla}}{{Nathanail} et~al.}{2021}]{Nathanail+21}
{Nathanail} A.,  {Gill} R.,  {Porth} O.,  {Fromm} C.~M.,   {Rezzolla} L.,
  2021, \mn@doi [\mnras] {10.1093/mnras/stab115}, \href
  {https://ui.adsabs.harvard.edu/abs/2021MNRAS.502.1843N} {502, 1843}

\bibitem[\protect\citeauthoryear{{Nava}, {Sironi}, {Ghisellini}, {Celotti}  \&
  {Ghirlanda}}{{Nava} et~al.}{2013}]{Nava+13}
{Nava} L.,  {Sironi} L.,  {Ghisellini} G.,  {Celotti} A.,   {Ghirlanda} G.,
  2013, \mn@doi [\mnras] {10.1093/mnras/stt872}, \href
  {https://ui.adsabs.harvard.edu/abs/2013MNRAS.433.2107N} {433, 2107}

\bibitem[\protect\citeauthoryear{{O'Connor} et~al.,}{{O'Connor}
  et~al.}{2023}]{OConner+23}
{O'Connor} B.,  et~al., 2023, \mn@doi [arXiv e-prints]
  {10.48550/arXiv.2302.07906}, \href
  {https://ui.adsabs.harvard.edu/abs/2023arXiv230207906O} {p. arXiv:2302.07906}

\bibitem[\protect\citeauthoryear{{Pe'er}}{{Pe'er}}{2012}]{Peer-12}
{Pe'er} A.,  2012, \mn@doi [\apjl] {10.1088/2041-8205/752/1/L8}, \href
  {https://ui.adsabs.harvard.edu/abs/2012ApJ...752L...8P} {752, L8}

\bibitem[\protect\citeauthoryear{{Piran}}{{Piran}}{1999}]{Piran-99}
{Piran} T.,  1999, \mn@doi [\physrep] {10.1016/S0370-1573(98)00127-6}, \href
  {https://ui.adsabs.harvard.edu/abs/1999PhR...314..575P} {314, 575}

\bibitem[\protect\citeauthoryear{{Racusin} et~al.,}{{Racusin}
  et~al.}{2008}]{Racusin+08}
{Racusin} J.~L.,  et~al., 2008, \mn@doi [\nat] {10.1038/nature07270}, \href
  {https://ui.adsabs.harvard.edu/abs/2008Natur.455..183R} {455, 183}

\bibitem[\protect\citeauthoryear{{Rees} \& {Meszaros}}{{Rees} \&
  {Meszaros}}{1992}]{Rees-Meszaros-92}
{Rees} M.~J.,  {Meszaros} P.,  1992, \mn@doi [\mnras]
  {10.1093/mnras/258.1.41P}, \href
  {https://ui.adsabs.harvard.edu/abs/1992MNRAS.258P..41R} {258, 41}

\bibitem[\protect\citeauthoryear{{Ren}, {Wang}, {Zhang}  \& {Dai}}{{Ren}
  et~al.}{2022}]{Ren+23}
{Ren} J.,  {Wang} Y.,  {Zhang} L.-L.,   {Dai} Z.-G.,  2022, arXiv:2210.10673,
  \href {https://ui.adsabs.harvard.edu/abs/2022arXiv221010673R} {}

\bibitem[\protect\citeauthoryear{{Ripa} et~al.,}{{Ripa} et~al.}{2023}]{Ripa+23}
{Ripa} J.,  et~al., 2023, \mn@doi [arXiv e-prints] {10.48550/arXiv.2302.10047},
  \href {https://ui.adsabs.harvard.edu/abs/2023arXiv230210047R} {p.
  arXiv:2302.10047}

\bibitem[\protect\citeauthoryear{{Rossi}, {Lazzati}  \& {Rees}}{{Rossi}
  et~al.}{2002}]{Rossi+02}
{Rossi} E.,  {Lazzati} D.,   {Rees} M.~J.,  2002, \mn@doi [\mnras]
  {10.1046/j.1365-8711.2002.05363.x}, \href
  {https://ui.adsabs.harvard.edu/abs/2002MNRAS.332..945R} {332, 945}

\bibitem[\protect\citeauthoryear{{Rossi}, {Lazzati}, {Salmonson}  \&
  {Ghisellini}}{{Rossi} et~al.}{2004}]{Rossi+04}
{Rossi} E.,  {Lazzati} D.,  {Salmonson} J.~D.,   {Ghisellini} G.,  2004,
  \mn@doi [\mnras] {10.1111/j.1365-2966.2004.08165.x}, \href
  {https://ui.adsabs.harvard.edu/abs/2004MNRAS.354...86R} {354, 86}

\bibitem[\protect\citeauthoryear{{Salafia} et~al.,}{{Salafia}
  et~al.}{2022}]{Salafia+22}
{Salafia} O.~S.,  et~al., 2022, \mn@doi [\apjl] {10.3847/2041-8213/ac6c28},
  \href {https://ui.adsabs.harvard.edu/abs/2022ApJ...931L..19S} {931, L19}

\bibitem[\protect\citeauthoryear{{Sari}}{{Sari}}{1997}]{Sari-97}
{Sari} R.,  1997, \mn@doi [\apjl] {10.1086/310957}, \href
  {https://ui.adsabs.harvard.edu/abs/1997ApJ...489L..37S} {489, L37}

\bibitem[\protect\citeauthoryear{{Sari} \& {Piran}}{{Sari} \&
  {Piran}}{1995}]{Sari-Piran-95}
{Sari} R.,  {Piran} T.,  1995, \mn@doi [\apjl] {10.1086/309835}, \href
  {https://ui.adsabs.harvard.edu/abs/1995ApJ...455L.143S} {455, L143}

\bibitem[\protect\citeauthoryear{{Sari} \& {Piran}}{{Sari} \&
  {Piran}}{1999a}]{Sari-Piran-99a}
{Sari} R.,  {Piran} T.,  1999a, \mn@doi [\apjl] {10.1086/312039}, \href
  {https://ui.adsabs.harvard.edu/abs/1999ApJ...517L.109S} {517, L109}

\bibitem[\protect\citeauthoryear{{Sari} \& {Piran}}{{Sari} \&
  {Piran}}{1999b}]{Sari-Piran-99b}
{Sari} R.,  {Piran} T.,  1999b, \mn@doi [\apj] {10.1086/307508}, \href
  {https://ui.adsabs.harvard.edu/abs/1999ApJ...520..641S} {520, 641}

\bibitem[\protect\citeauthoryear{{Sato}, {Murase}, {Ohira}  \&
  {Yamazaki}}{{Sato} et~al.}{2023}]{Sato+23}
{Sato} Y.,  {Murase} K.,  {Ohira} Y.,   {Yamazaki} R.,  2023, \mn@doi [\mnras]
  {10.1093/mnrasl/slad038}, \href
  {https://ui.adsabs.harvard.edu/abs/2023MNRAS.tmpL..35S} {}

\bibitem[\protect\citeauthoryear{{Sironi} \& {Spitkovsky}}{{Sironi} \&
  {Spitkovsky}}{2011}]{Sironi-Spitkovsky-11}
{Sironi} L.,  {Spitkovsky} A.,  2011, \mn@doi [\apj]
  {10.1088/0004-637X/726/2/75}, \href
  {https://ui.adsabs.harvard.edu/abs/2011ApJ...726...75S} {726, 75}

\bibitem[\protect\citeauthoryear{{Troja} et~al.,}{{Troja}
  et~al.}{2019}]{Troja+19}
{Troja} E.,  et~al., 2019, \mn@doi [\mnras] {10.1093/mnras/stz2248}, \href
  {https://ui.adsabs.harvard.edu/abs/2019MNRAS.489.1919T} {489, 1919}

\bibitem[\protect\citeauthoryear{{Williams} et~al.,}{{Williams}
  et~al.}{2023}]{Williams+23}
{Williams} M.~A.,  et~al., 2023, \mn@doi [arXiv e-prints]
  {10.48550/arXiv.2302.03642}, \href
  {https://ui.adsabs.harvard.edu/abs/2023arXiv230203642W} {p. arXiv:2302.03642}

\bibitem[\protect\citeauthoryear{{Yan}, {Wei}  \& {Fan}}{{Yan}
  et~al.}{2007}]{Yan+07}
{Yan} T.,  {Wei} D.-M.,   {Fan} Y.-Z.,  2007, \mn@doi [\cjaa]
  {10.1088/1009-9271/7/6/05}, \href
  {https://ui.adsabs.harvard.edu/abs/2007ChJAA...7..777Y} {7, 777}

\bibitem[\protect\citeauthoryear{{Zhang}, {Liu}, {Geng}, {Wu}  \&
  {Wang}}{{Zhang} et~al.}{2022}]{Zhang+22}
{Zhang} Z.-L.,  {Liu} R.-Y.,  {Geng} J.-J.,  {Wu} X.-F.,   {Wang} X.-Y.,  2022,
  \mn@doi [\mnras] {10.1093/mnras/stac1198}, \href
  {https://ui.adsabs.harvard.edu/abs/2022MNRAS.513.4887Z} {513, 4887}

\bibitem[\protect\citeauthoryear{{de Ugarte Postigo} et~al.,}{{de Ugarte
  Postigo} et~al.}{2022}]{deUgartePostigo+22}
{de Ugarte Postigo} A.,  et~al., 2022, GRB Coordinates Network, \href
  {https://ui.adsabs.harvard.edu/abs/2022GCN.32648....1D} {32648, 1}

\makeatother
\end{thebibliography}







\bsp	
\label{lastpage}
\end{document}